\documentclass[preprint]{revtex4-1}
\usepackage{graphicx}
\usepackage{pstricks}
\usepackage{amsmath}
\usepackage{bm}

\begin{document}

\title{Electronic, optical and charge transfer properties of topologically protected states in hybrid bismuthene layers}
\author{Erika Nascimento Lima}
\affiliation{Universidade Federal do Mato Grosso, Campus Rondon\'opolis, Rondon\'opolis, Mato Grosso,  Brazil}
\author{Andreia Luisa da Rosa (ORCID: 0000-0002-2780-6448)}
\affiliation{Universidade Federal de Goi\'as, Institute of Physics, 74690-900 Goi\^ania, Goi\'as, Brazil}
\affiliation{Bremen Center for Computational Materials Science, University of Bremen, Am Fallturm 1, 28359 Bremen, Germany}
\author{Renato Borges Pontes (ORCID: 0000-0002-3336-3882)}
\affiliation{Universidade Federal de Goi\'as, Institute of Physics, 74690-900 Goi\^ania, Goi\'as, Brazil}
\author{Tome Mauro Schmidt}
\affiliation{Universidade Federal de Uberl\^andia, Institute of Physics, 38400-902 Uberl\^andia, Minas Gerais, Brazil}
\author{Jailton Almeida}
\affiliation{Universidade Federal da Bahia, Institute of Physics, 40210-340 Salvador, Bahia, Brazil}

\author{Thomas Frauenheim (ORCID: 0000-0002-3073-0616)}
\affiliation{Bremen Center for Computational Materials Science, University of Bremen, Am Fallturm 1, 28359 Bremen, Germany}
\author{Maur\'icio Chagas da Silva (ORCID: 0000-0002-6890-0182)}
\affiliation{Max-Planck-Institute for the Structure and Dynamics of Matter, Luruper Chaussee 149,  22761 Hamburg, Germany}
\affiliation{Bremen Center for Computational Materials Science, University of Bremen, Am Fallturm 1, 28359 Bremen, Germany}

\begin{abstract}

  We have performed first-principles calculations of electronic and
  dielectric properties of single-layer bismuth (bismuthene) adsorbed
  with -COOH. We show that the Bi-COOH hybrid structure is a
  two-dimensional topological insulator with protected edge Dirac
  states. The adsorption process of -COOH induces a
  planar configuration to the initially pristine buckled
  bismuthene. We claim that the stability of these planar structure
  mainly stem from strain induced by the adsorption of the -COOH
  organic group, but it is also related to ligand-ligand
  interactions. Using charge density analysis we show that the role of
  this organic group  is not only to stabilize the layer but
  also to functionalize it, which is very important
  for future applications such as sensing, biomolecules
  immobilization, as well in electronic spintronic 
  and even optical devices, due to its large band gap. 
  Finally we demonstrate that many body corrections
  are crucial to obtain a better description of the electronic and
  dielectric properties of these systems.

\end{abstract}


\maketitle

\section{Introduction}

Bismuth single layers were theoretically predicted to have topological
insulator behavior with protected edge non-trivial Dirac states
\,\cite{Kou:NL,Rivelino2015,AsiaNat}, which are robust against 
perturbations that preserve time reversal symmetry, like strain
and external electric
field\,\cite{MingYang2017}.  Only recently the existence of pure,
ultra-thin bismuth layers has been experimentally confirmed, which
were synthesized on a silicon carbide
substrate\,\cite{Reis2017}. Although bismuthene has already inherent
properties such as topological non-trivial band structure, it has been
proposed that these bismuth layers are very sensitive to changes
in the environment \cite{Kou2018a,Rivelino2015}. 

A promising route for tuning the electronic properties of bare layered materials
is the adsorption of organic functional groups
\,\cite{ProgressReports}.  Previously, -H and ${\rm -CH_3}$ groups
have been investigated. Both groups were predicted to preserve the
topological insulator behavior of
bismuthene\,\cite{Kou:NL,Rivelino2015}. Furthermore, valley-polarized
quantum anomalous Hall states emerge as these bismuth layers films are
adsorbed with hydrogen\,\cite{Niu2015}, enlarging the band gap, indicating that 
ligand-mediated interactions with the substrate may be a route to functionalize this 2D TI. 
More recently it has been  proposed that -COOH groups can induce ferroelectric behavior
on bismuthene layers\,\cite{Kou2018a}. Theoretical calculations of
optical properties of a single-layer bismuthene have been recently
reported by Kecik {\it et al.} \cite{Ciraci2019}. In such investigation, an exciton
with binding energy of 0.18 eV has been found, with a strong absorption 
peak around 2.2 eV, indicating that such 2D material can be useful for 
applications, in solar cells, light-emitting devices, photodetectors, among others.

Despite all that, the role of organic ligands on bismuthene on
introducing new functionalities, such as change of reactivity, which
is a requirement for further applications such as immobilization of
organic or biomolecules and also in the field of catalysis, has been
little discussed. Furthermore, dielectric and optical properties of
hybrid bismuthene-organic systems remain unexplored.

In this work, we employ density-functional theory (DFT) and 
many-body theory (GW) to
show that the bismuthene-based hybrid structures preserve the topological insulating
behavior. We show that the adsorption of -COOH groups induces a planar
geometry on the initially buckled bismuthene structure. Most
importantly our results reveal that the role of -COOH groups is not only to
protect the 2D layer but also to functionalize the
bismuth layers by changing the reactivity of the hybrid system.
Furthermore we perform GW calculations to obtain the electronic band
gap and dielectric properties of the hybrid layers. We find that
self-consistency in the GW calculations is necessary to obtain a
better description of the dielectric properties compared to
single-shot GW calculations. Finally, we suggest that the charge
transfer between the ligand and the bismuth layer promoted by
adsorption is important for future applications such as sensing and
biomolecules immobilization.

\section{Methodology}

  The electronic structure and the chemical bond analysis of
  bismuthene adsorbed with -COOH groups were investigated within
  density functional theory framework. The projected augmented wave method
  (PAW)\,\cite{Bloechel:94,Kresse:99} combined with the generalized
  gradient approximation\,\cite{Perdew:96} as implemented in the VASP
  package has been employed\,\cite{Kresse:99,Shishkin:07}. The lattice
  parameters and atomic position were fully relaxed until the atomic
  forces converge to less than 10$^{-4}$\,eV/{\AA}. A vacuum region of
  at least 10\,{\AA} perpendicular to the bismuth surface was
  introduced to avoid interaction between atoms in neighboring cells.
  The reciprocal space was sampled with a (10$\times$10$\times$1) {\bf
    k}-point $\Gamma$-centered grids to relax the structures. The wave
  functions were expanded in a plane-wave basis set with an energy
  cutoff of 500\,eV. Spin orbit coupling (SOC) was included in all
  calculations, including relaxation, electronic structure and
  dielectric properties. The dielectric function was calculated using
  the GW approximation with 8 iterations in the Green's function G,
  which was found to be suficient enough to converge the
  quasi-particle eigenvalues with respect to G$_0$W$_0$ calculations.
  In the later case, a $(5\times5\times1)$ {\bf k}-point grid was used
  due to the high computational costs. Tests with up to a
  $(12\times12\times1)$ {\bf k}-points for bare single-layered bismuth
  have shown that this smaller set is enough to achieve
  convergence. Moreover, due to high computational cost 
  the energy cutoff for GW calculations has been
  decreased to 400\,eV.
       
\section{Results and Discussion}

The stability of the pristine buckled and the planar bismuthene has been
investigated by varying the lattice parameters $a$ and $b$ and fully
relaxing all the atoms in the unit cell. The optimized in-plane
lattice parameters are $a$ = $b$ = 4.39 \AA\ for the buckled structure
and $a$ = $b$ = 5.30 \AA\ for the planar one. Our results are in good
agreement with the results obtained by Freitas et al. \cite{Rivelino2015}, whose DFT-based calculations indicate
 that buckled bismuthene exhibits an equilibrium structure with a buckling parameter $\Delta$ = 1.73 \AA\ and a lattice constant $a$ = 4.34 \AA. For a planar bismuthene they obtained a lattice constant $a$= 5.57 \AA.
Our calculated structural parameters for bulckled bismuthene are also in very good agreement with previous theoretical and experimental values in the literature \cite{Rasche2010,Hirahara2012,Cheng2014,Pumera2017,Liu2019}.

The inclusion of SOC does not change significantly the lattice
parameters, but it lowers the total energy of the system. We found
that the buckled structure is much more stable than the planar
configuration with an energy difference of 1.55 eV. The electronic
stability of the buckled structure compared to the planar one can be
understood considering that in two-dimensional bismuthene each bismuth
atom has three nearest neighbors forming $\sigma$-like bonds. In order
to achieve the electronic stability, the $sp^3$ hybridization is
favored in respect to the the planar $sp^2$-type hybridization,
leading to a buckled structure. This structural configuration can be
explained by the qualitative assumptions of the valence shell electron
pair repulsion (VSEPR) theory\cite{VSEPR}. The VSEPR theory suggests
that the electronic density of molecules, localized bonds and/or lone
electrons pairs, seeks for spatial configuration which minimizes the
electron-electron interactions, i.e., the angles between the localized
bonds and/or lone electrons pair should be greatest to minimize the
repulsion between them in a such spatial arrangement. Thus, the $sp^3$
hybridization presents a angle of about 109$^{\circ}$ between the lone
electrons pairs of Bi centers and the Bi-Bi $\sigma$ bonds. This
geometric arrangement has an angle which is larger than the
90$^{\circ}$ considering a $sp^2$ hybridization.  This minimizes the
electron-electron interactions between the Bi-Bi $\sigma$-like bond
and the non-bonded lone electrons pairs of Bi centers producing a
buckling of 1.72\,{\AA}. In this configuration, bismuthene has a band
gap of 0.9\,eV, in good agreement with previously reported
results\,\cite{Kou2018a,Rivelino2015}.

In this investigation, we have adsorbed -COOH  on bismuthene at one
monolayer (ML) coverage regime. The ligands are adsorbed 
on top positions of bismuth atoms on both sides of the
bismuth surface. We verified that the configurations where the
radicals are on top positions are more stable than hollow or bridge
sites. Although, we cannot rule out that smaller coverages may be
present, we should point out that lower coverages were found to yield
less stable structures in other two-dimensional hybrid systems, such
as germanene\,\cite{JiangNT:2014}.

\begin{figure}[ht!]
\includegraphics[width=8cm,scale=1.0,clip]{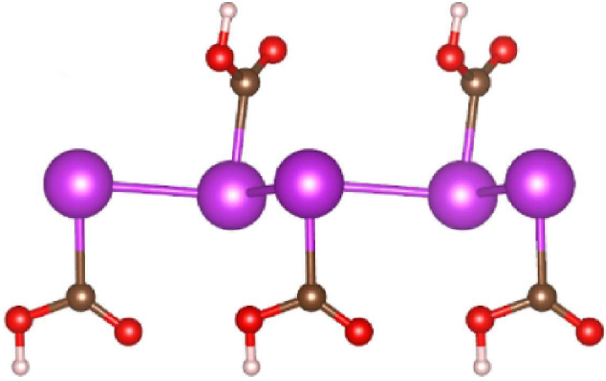}
\caption{\label{fig:structures}(color online) Side view of bismuth layer 
adsorbed with ${\rm -COOH}$ at 1 ML coverage regime.} 
\end{figure}

Fig.\,\ref{fig:structures} shows a side view of relaxed hybrid
bismuthene-COOH layer. Upon adsorption of -COOH group, the initial
buckled geometry of bismuthene relaxes to a near planar structure,
with an in-plane lattice parameter of 5.10\,{\AA}.  There are two main
driving forces for the planarity of these structures: the
ligand-ligand interaction and the electronegativity of the
ligand. Again, according to the VSERP theory\,\cite{VSEPR}, the $sp^3$
hybridization should be preferred in the adsorbed layers against the
$sp^2$ in bare bismuthene. However, lateral ligand-ligand interactions
play an important role as well. As a consequence of these additional
ligand-ligand interactions, the system assumes a quasi-planar
configuration which increases the in-planar lattice constant. The
hybrid system then assumes a quasi-planar arrangement in which the
Bi-Bi bonds are larger compared to pure bismuthene and Bi-Bi-C angle
is reduced to almost 90$^{\circ}$ in a $sp^3$ hybridization, as shown in Fig.\,\ref{fig:structures}. A similar behaviour has been suggested
for organic functionalized germanene\,\cite{JiangNT:2014}.

\begin{figure}[ht!]
\begin{center}
\begin{tabular}{c}
  \includegraphics[width = 12cm,scale=1, clip = true]{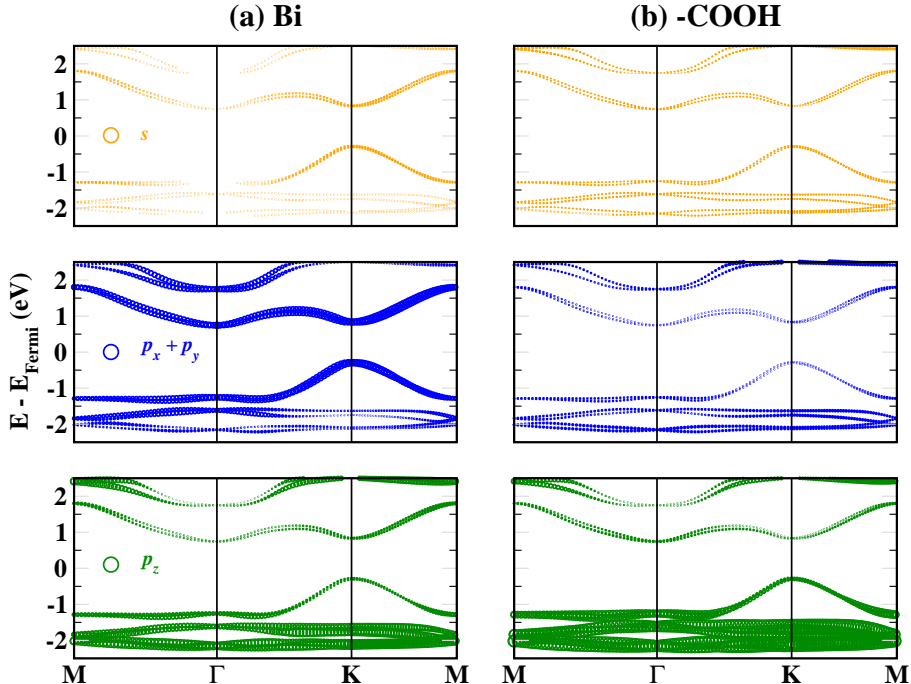}
\end{tabular}
\end{center}
\caption{\label{fig:projected_bands} Projected band structures on $p_x$ (blue), 
$p_y$ (red) and $p_z$ (orange) orbitals. Without the inclusion of SOC project: a) on bismuth 
and b) on ligand atoms. And with the inclusion of SOC projected on c) on bismuth and d) on 
ligand atoms. The Fermi level is set to zero energy in all plots.}
\end{figure}

In the absence of SOC, Bi-COOH exhibits semi-metallic character and fourfold
degeneracy at the $\Gamma$ point. From the orbital projected band
structure in Fig.\,\ref{fig:projected_bands} (a) and (b), one can see
that the $p_x$, $p_y$ and $p_z$ orbitals dominate the bands near the
Fermi level, mainly coming from Bi atoms. The change in the geometry and
inclusion of SOC upon -COOH adsorption shown in
Fig.\,\ref{fig:projected_bands} (c) and (d) shifts the conduction band
minimum (CBM) from K to $\Gamma$. We can see some contributions
from the -COOH group at valence band maximum (VBM) and conduction band
minimum (CBM). An indirect band gap is found since the VBM is at the K
point and the CBM lies at the $\Gamma$ point. The resulting indirect band gap of 1.0 eV is 10$\%$ 
larger than the bare bismuthene band gap. It is also possible to identify further transitions of 2.0 eV
at $\Gamma$, 3.2 eV at M and 1.1 eV at K points. This is similar to
what has been reported in H and CH$_3$ adsorbed on bismuth
layers\,\cite{Kou:NL,Rivelino2015,unpublished} using GGA
calculations. One can see that most ${\rm p_x}$ and ${\rm p_y}$
orbitals belonging to bismuth contribute to states close to Fermi
level. Molecular states also lie close to Fermi level.  The CBM
contains both ${\rm p_x}$ and ${\rm p_y}$ characters of bismuth and
ligand. As a conclusion, the hybrid systems contain significant
contribution from both bismuth and ligand orbitals. However, the
interaction between ligands at such a high coverage induces strain,
which weakens the Bi-Bi bonds, as discussed above. Although this
effect makes the planar bonds larger, the bismuth-ligand interaction is
strong enough to stabilize the hybrid layers. This will be
discussed below, as there is a net charge accumulation of electrons on
the ligand oxygen atom.  We should mention that electrons $s$ do not
contribute to states close to the Fermi level, since they appear far
below in energy from the top of the valence band.

\begin{figure}[ht!]
\includegraphics[width = 12cm,scale=1, clip = true]{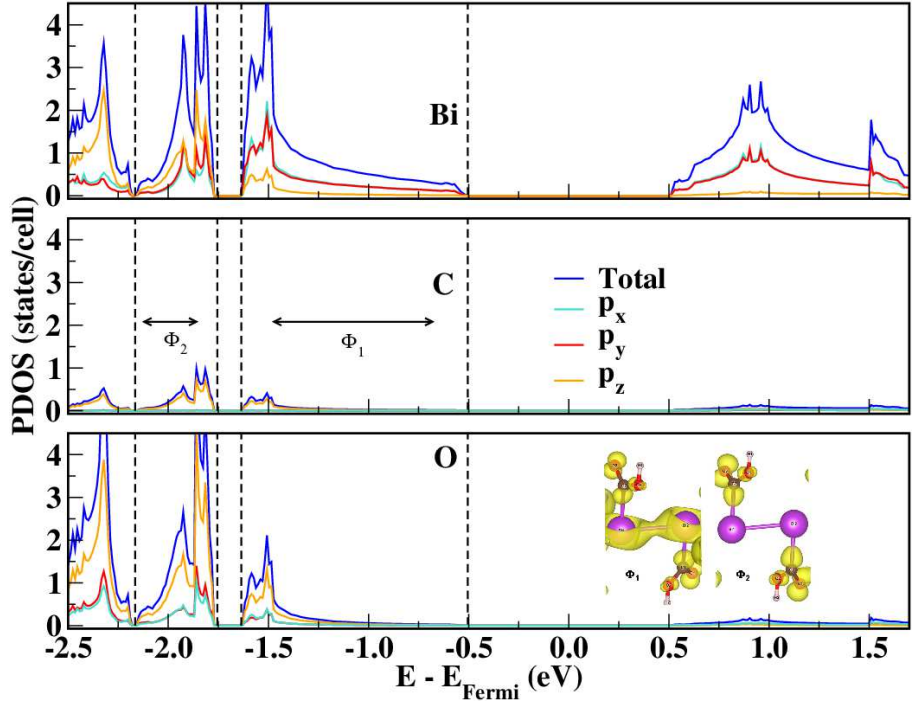}
\caption{\label{fig:PDOS} Total charge density showing the regions $\Phi_1$  and $\Phi_2$ 
and orbital resolved projected density of states for monolayer Bi-COOH.
The Fermi level is set at zero energy in all plots. 
The insets stand for charge density plots 
of the regions $\Phi_1$  and $\Phi_2$, respectively. The Fermi level is set at zero.}
\end{figure}

In addition we show the projected density of states (PDOS) on the
individual atoms in Fig. \ref{fig:PDOS}. The atomic orbital
interpretation of the bands close to the Fermi level can be split
into two main energetic range regions, namely $\Phi_1$ and $\Phi_2$,
shown in the inset of Fig. \ref{fig:PDOS} (c).  We can see that the
contribution of the carbon orbitals are very small compared to the
bismuth and oxygen atoms in both $\Phi_1$ and $\Phi_2$
regions. Moreover, the $\sigma$-like bond between Bi-C has a large
contribution of the bismuth $p_z$ orbitals with a large electron
donation to carbons atoms. In this scenario, the bismuth
atoms act as a Lewis basis donating electronic density through the $\sigma$ bond. 
On the other hand, the carbon atoms act as a Lewis acid receiving electronic density
from the bismuth atoms via the $\sigma$ bond.

The Bi-C $\sigma$-like bond is better represented by the $\Phi_2$
energetic band region shown in Fig. \ref{fig:PDOS}. The $\Phi_1$
region is majority Bi-Bi bonds which can be interpretated as a
combination of the in-plane orbitals, $p_x$ and $p_y$ orbitals. The
$\Phi_1$ region contemplates the top of the VB region, and so 
it dictates the band gap behavior. As we can see from the charge 
density in Fig. 3,  $\Phi_1$ comes mostly from the in-plane Bi-Bi ($p_x$,$p_y$) bonds. 
So, the increase in the band gap of the Bi-COOH with respect to the bare 
bismuthene must be due to the flattening Bi-Bi bonds. 
The $\Phi_2$ region is mostly associated to Bi-C $\sigma$-like
bond as it can be seen in Fig. \ref{fig:PDOS}. The main frontier bands
of Bi-COOH is basically formed by Bi-Bi bonds with delocalized
orbitals, which can be explained by the composition of $p_x$ and $p_y$
Bi orbitals. Moreover, the Bi-C bond lies also deeper inside the valence band
demonstrating in fact the presence of a Bi-C bond.

\begin{figure}[ht!]
  \includegraphics[width = 9cm, scale=1.1, clip = true ]{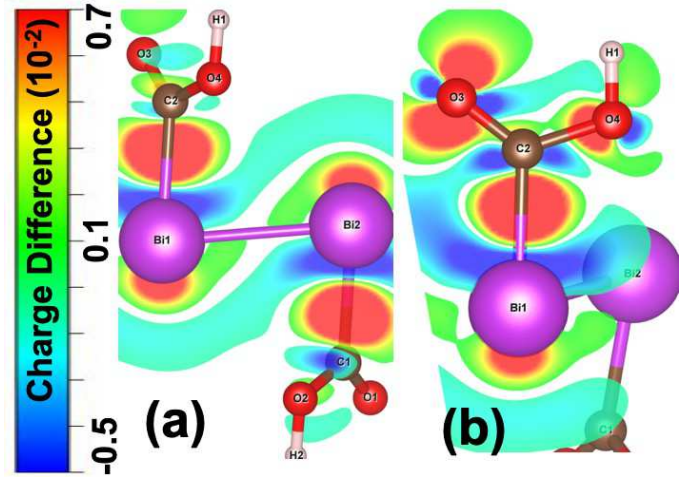}
\caption{Two views of the charge density difference ($\Delta\rho$) for monolayer 
Bi-COOH. Negative (positive) values represent regions where electrons were accumulated 
 (withdrawn) after adsorption.} 
\label{fig:drho}
\end{figure}

The nature of the chemical bonds in the hybrid system was
investigated by the charge density difference between the electronic
densities of the Bi-COOH and their constituent systems. The electronic
density difference ($\Delta\rho$) is given by: ${\rm \Delta\rho = \rho^{Bi-COOH}
  -\rho^{Bi} - \rho^{-COOH}}$, where ${\rm \rho^{Bi-COOH}}$ is the
electronic density of the hybrid Bi-COOH layers, $\rho^{Bi}$ and
${\rho^{-COOH}}$ are the charge densities calculated at fixed atomic
positions of the Bi and -COOH, respectively. The charge density difference
shown in Fig. \ref{fig:drho} reflects on how the electronic charge density
changes upon -COOH adsorption. The complex Bi-COOH has a high
electronic density between Bi-C atoms. It is also seen that the
electronic charge density accumulates half-way of the bond regions,
decreasing the electronic density around the atomic regions (shown in
blue). The Bi-Bi bond suffer small changes, while a strong Bi-C bond can be seen in Fig. 4-b. 
Since upon cleavage of the bulk (111) surface the bismuth
atoms are left with two extra electrons on the unsaturated bonds, upon
-COOH adsorption covalent bonds between C and Bi bonds are expect to
form. We can also notice that the charge is further distributed to the
oxygen atoms on the ligand (O1 and O3). This is an important conclusion,
since it reveals that the role of -COOH is not only to protect the
Bi surface, but also to functionalize the bismuth layers.

\begin{figure}[ht!]
\includegraphics[width = 9cm,scale=1, keepaspectratio, clip = true]{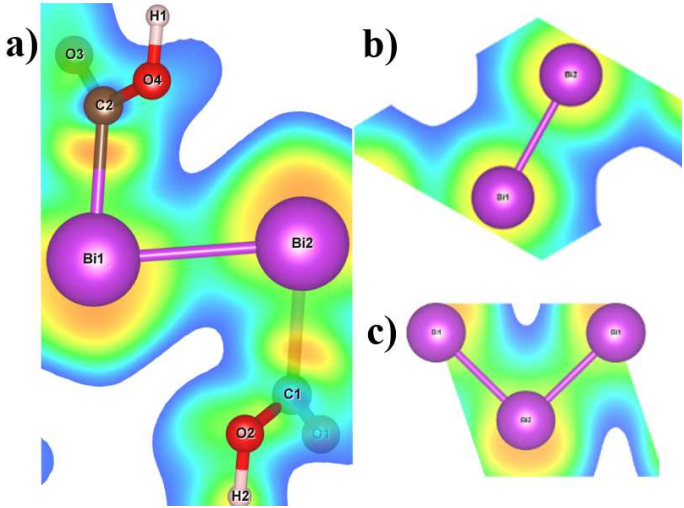} 
\caption{(a-c) Different views of the electronic localization function cross-section
 for a single-layer Bi-COOH.  The metallic feature of bismuth bonds is shown in
 green. The localized regions due to covalent bonds and lone-pair
  electrons of non-bonding orbitals are shown in orange.}
\label{fig:elf}
\end{figure}

In order to have insight on the nature of the chemical bond
environment in the functionalized Bi-COOH system, the electronic
localization function (ELF)\,\cite{relf} was also evaluated. ELF
represents the electronic density localization in the spatial region
of crystals and molecules, whose values vary between 0 to 1.  ELF
values about 0.5 can be interpreted as metallic type bonds (green in
Fig. 5), whereas values close to 1.0 indicate localized electronic
densities (covalent bonds between atoms or a non-bonding lone electron
pairs, shown in red color in Fig. 5)).  The ELF of the Bi-COOH,
Bi-buckled and Bi-planar are presented in Fig.\,\ref{fig:elf}.  In
Fig.\,\ref{fig:elf}(a), a region of localized electronic density
between carbon and bismuth atoms can be clearly identified (ELF values
close to 1.0), thus characterizing a covalent bond. In addition, it is
possible to see that the C-Bi chemical bond has a slight ionic
characteristic which is usual for -COOH groups. The low electronic
charge density between the bismuth atoms is also verified for the
Bi-planar structure, shown in Fig.\,\ref{fig:elf}(c). However, in the
Bi-buckled structure, the ELF reveals a metallic bond nature between
Bi bonds, \ref{fig:elf}(b). The formation of the Bi-COOH bonds induced
a distortion into the geometry from a chairlike configuration in
Bi-buckled to a planar configuration. The chemical behavior of -COOH
group withdraws electronic density of the Bi atoms, which it
contributes to the reduction of the electronic density along Bi-Bi
bonds. We can then conclude that the adosrption of -COOH groups change
not only the geometrical feature of bismuthene, but also introduces a
charge transfer between the ligand and the bare bismuth layers and
therefore a change in the reactivity.

\begin{figure}[ht!]
\includegraphics[width = 16cm,scale=1, clip = true]{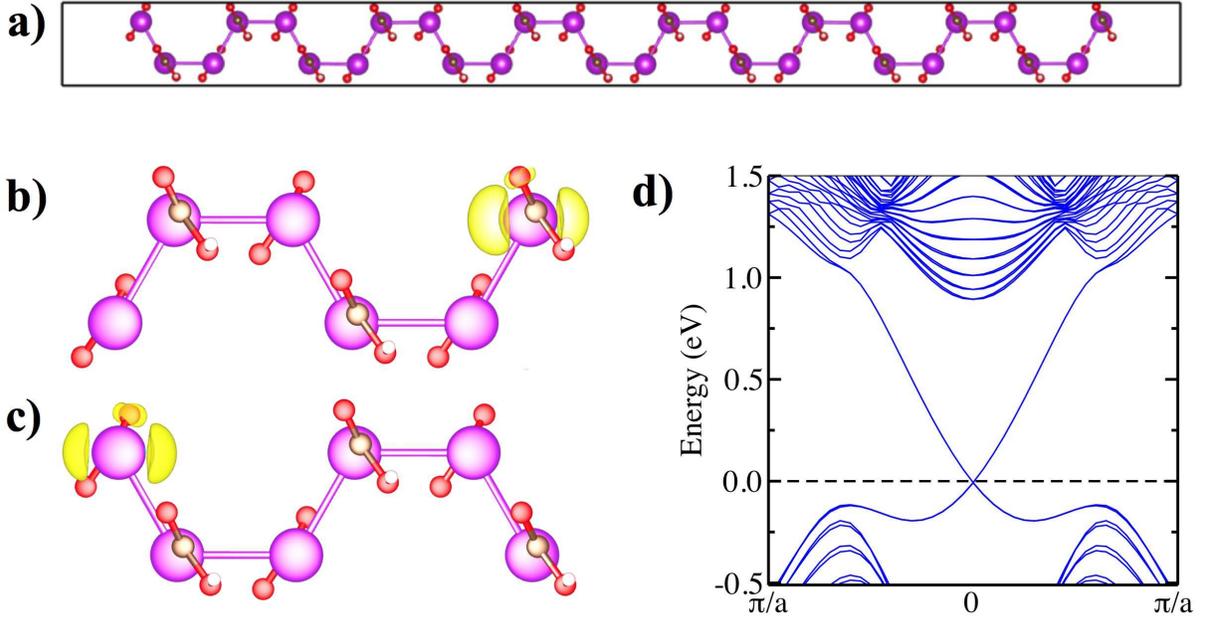}
\caption{a) DFT-GGA relaxed geometry of the Bi-COOH nanoribbon, 
b) band decomposed charge density at
  $\Gamma$ point (spin up), c) band decomposed charge density at
  $\Gamma$ point (spin down) and d) electronic band structure of a
  Bi-COOH nanoribbon.  The optimized in-plane lattice parameters are:
  $a$= 5.5 \AA\ and $b$= 63.55 \AA. The Fermi level is set at
  zero. Isosurface value is set to ${\rm 0.002 \;e/{\AA}^3}$.}
\label{fig:ribbon}
\end{figure}

\begin{figure}[ht!]
\includegraphics[width = 16cm,scale=1, clip = true]{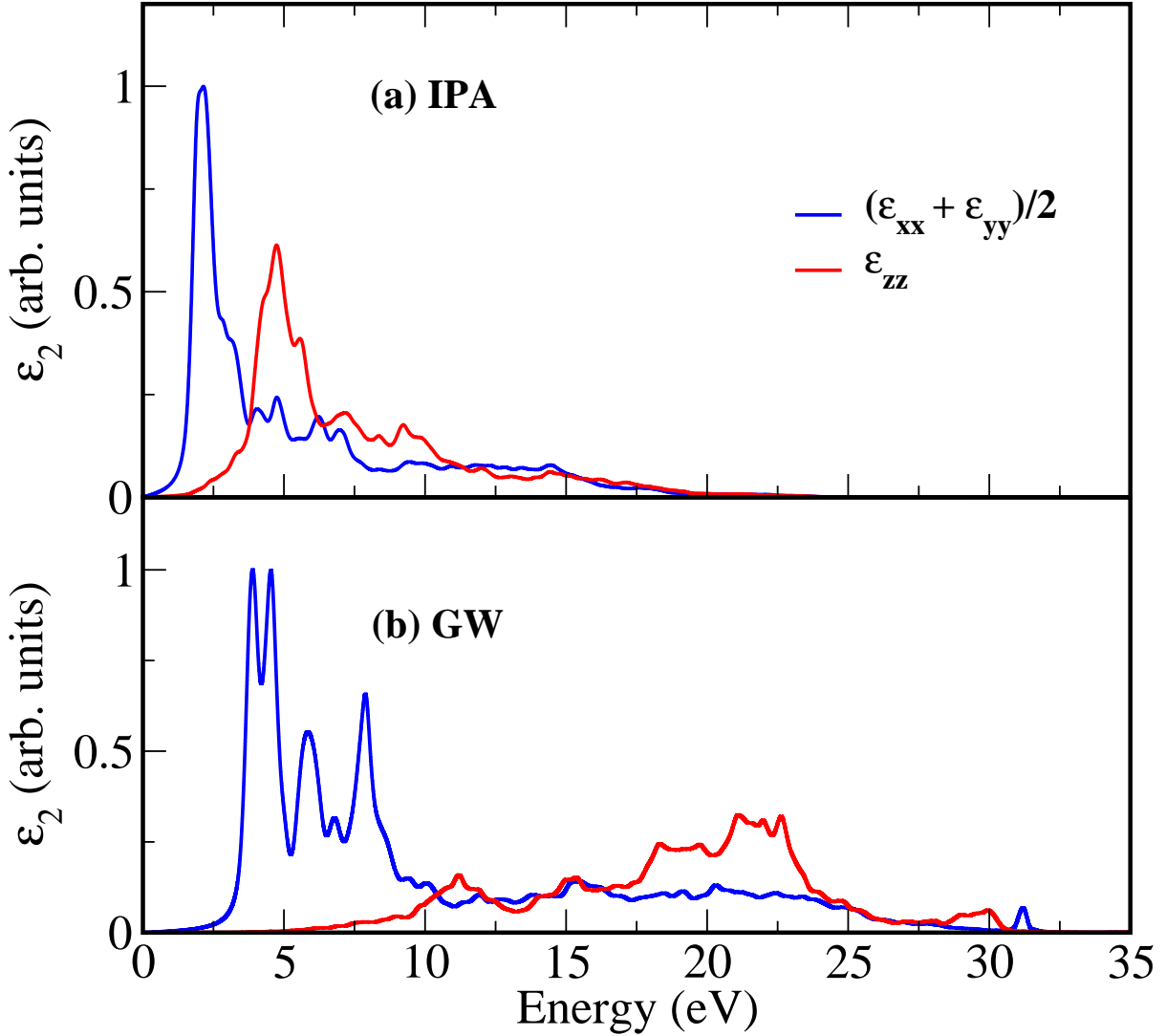}
\caption{Imaginary electronic dielectric function ($\varepsilon_2$) 
profile, for Bi-COOH system, in the:
(a) independent particle approximation (IPA) and (b) GW approximation.}
\label{fig:IPA-GW}
\end{figure}

The question that rises now is whether the hybrid system have a
topologically nontrivial band structure. We have calculated the Z$_2$
topological invariant according to xx. This number corresponds to the
number of Kramers pairs of edge modes, integrating over half of the
Brillouin zone. If the overall Z$_2$ sum of occupied bands is even,
the system is a regular insulator. On the other hand, if the sum is
odd, the system is a topological insulator. We find that Z$_2$ is
equal to 1, which means that the hybrid Bi-COOH layres behave as
topological insulator.  Additionally, we have calculated the band
structure of a bismuth nanoribbon with zig-zag edges adsorbed with
-COOH, as shown in Fig.  \ref{fig:ribbon} (a). Once the edges are
exposed, it is possible to verify whether the topological states come
from the edge atoms. The Bi-COOH nanoribbons contain 28 Bi atoms, 28 C
atoms, 56 O atoms and 28 H atoms and have width of 60\,{\AA}, which
suffices to avoid spurious interaction between two edge-ends (a ribbon
with 102\,{\AA} width leads to identical
results). Figs.\,\ref{fig:ribbon} (a) and (b) shows the band projected
electronic charge density at the $\Gamma$ point for the edge
states. It is clearly seen that the spin up, Fig.\,\ref{fig:ribbon}
(b) and spin down Fig.\,\ref{fig:ribbon} (c) states are very localized
at the edges only, confirming a non-trivial character of the band
structure.  The calculated electronic band structure of the
functionalized nanoribbon is shown in Fig.\ref{fig:ribbon} (d). One
can see the topological edge states with a single Dirac crossing at M
points, confirming that the edge states are protected by time reversal
symmetry.  Since the band gap of the hybrid Bi-COOH is enlarged with
respect to bare bismuthene, optical applications can be important in
this 2D TI material. Besides that Bi 5d orbital is completely filled,
and many-body effects are expected to be significant [9].  Thus, we
have calculated the dielectric function ($\varepsilon$).  In
particular, we computed the imaginary term of the electronic
dielectric function ($\varepsilon_2$) for the Bi-COOH hybrid system
within both the independent particle (IPA) and GW approximations.  The
$\varepsilon_2$ is shown in Fig.\ref{fig:IPA-GW}. The inclusion of SOC
in the optical properties is also taken into account. Both IPA,
Fig.\ref{fig:IPA-GW}(a), and GW, Fig.\ref{fig:IPA-GW}(b),
approximations show an anisotropic behaviour. In the IPA, an intense
absorption band around 2.1\,eV is seen for the averaged
($\varepsilon_2^{xx}$ + $\varepsilon_2^{yy}$)/2 parallel component.
On the other hand, the $\varepsilon_2^{zz}$ component is broadly
distributed in the region 0-10 eV. In the IPA, the electronic
(indirect) band gap is 1.0 eV, which correspond to the onset seen
Fig.\,\ref{fig:IPA-GW}(a). The inclusion of many-body effects in the
G$_0$W$_0$ approximation, i.e., without self-consistency neither in
the Green's function G nor in the screened Coulomb potential W,
improves the electronic band gap a little to 1.01\,eV.  However,
self-consistency in G gives a value of 2.4\,eV to the band gap of the
hybrid Bi-COOH, which is larger than the G$_0$W$_0$ value.  This
results is shown in Fig.\ref{fig:IPA-GW}(b).  The first strong
transition is around 3\,eV. A large anisotropy is also seen.  The
$\varepsilon_2^{zz}$ component is a broad function between 5 and
30\,eV. Our results show that the inclusion of many-body effects are
important to obtain a better description of the dielectric properties.

\section{Conclusions}

We have performed density-functional theory and GW calculations of
electronic and dielectric properties for bismuthene adsorbed with
-COOH groups.  The electronic properties of this hybrid system show
that the functionalized layers have topological insulating behavior
with a reasonable band gap. Our results reveal that the stability of these planar
structure stem from both reactivity change induced by the adsorption
of -COOH on the buckled bismuth layers promoting C-Bi
bonds. Furthermore, the role of -COOH groups is to functionalize the
bismuth layers, since oxygen atoms belonging to the ligand become
reactive. This reactivity is desired for enhanced catalysis or
immobilization of organic and biomolecules. Finally we show that
inclusion of many-body effects is very important to obtain a better
description of the dielectric properties of the hybrid system compared
to the independent particle approximation.

\section{Acknowledgements}

We acknowledge the financial support from the Brazilian Agencies CNPq
and FAPEG (PRONEX 201710267000503) and German Science Foundation (DFG)
under the program FOR1616. The calculations have been performed using
the computational facilities of CENAPAD-SP and LNCC (Supercomputer Santos Dumont) 
and at QM3 cluster at the Bremen Center for Computational Materials Science.

\bibliographystyle{apsrev}

\end{document}